\documentstyle[12pt]{article}

\newcommand{\alt}{\mathrel{\mathpalette\vereq<}}
\def\vereq#1#2{\lower3pt\vbox{\baselineskip1.5pt \lineskip1.5pt
\ialign{$#1\hfill##\hfil$\crcr#2\crcr\sim\crcr}}}
\renewcommand{\theequation}{{\protect\thesection.\arabic{equation}}}
\begin{document}

\title{Random walks and the correlation length critical exponent\\
 in scalar quantum field theory\footnote{Submitted to Physical 
Review D.}\\ [.25in]}
\author{Joe Kiskis\\ [.7ex]
Department of Physics\\
University of California\\
Davis, CA 95616, USA\\ 
{\tt jekucd@ucdhep.ucdavis.edu}\\[.15in]
Rajamani Narayanan\\ [.7ex]
Theoretical Division\\
DESY\\
D-W-2000 Hamburg 52, Germany\\[.15in]
Pavlos Vranas\\ [.7ex]
Supercomputer Computations Research Institute\\
 Florida State University\\
Tallahassee, FL 32306, USA}

\maketitle
\newpage
\begin{abstract}
The distance scale for a quantum field theory is
the correlation length $\xi$, which diverges with exponent $\nu$ as the bare
mass approaches a critical value. If $t=m^{2}-m_{c}^{2}$, then
$\xi=m_{P}^{-1} \sim t^{-\nu}$ as $t \rightarrow 0$.
The two-point function
of a scalar field has a random walk
representation. The walk takes place in a background of fluctuations (closed
walks) of the field itself. We describe the connection between properties of
the walk and of the two-point function. Using the known behavior of the two
point function, we deduce that the dimension of the walk is
$d_{w}=\varphi / \nu$
and that there is a singular relation between $t$ and the energy per unit
length of the walk $\theta \sim t^{\varphi}$ that is due to the singular
behavior of the background at $t=0$. ($\varphi$ is a computable crossover
exponent.)
\end{abstract}
\vspace{3in}
\newpage

\section{Introduction}
\setcounter{equation}{0}

The two-point function of a scalar quantum field has a random walk
representation. The divergence of the correlation length has a critical
exponent $\nu$. This paper elucidates the relation between properties of the
walk and the value of the exponent.

The exponent $\nu$ can be approximately calculated in quantum field theory
using methods such as the $\epsilon$-expansion, the strong coupling expansion,
and Monte Carlo numerical simulations. In spite of the successes of these
calculations, they do not provide much physical insight beyond the general
ideas of scaling and the renormalization group. Scalar quantum field theory
can be given a random walk representation \cite{r1}.
The associated physical picture is
concrete and intuitive. It provides some helpful insights.

A tremendous amount of work has been done on random walks outside of the context
of quantum field theory. Let $P(x,l)$ be the probability that a walk of length
$l$ ends at $x$.
The large-distance behavior is characterized by a scaling form
\cite{r3}
\begin{equation}
   P(x,l)=p(x/R(l))
\end{equation}
with
\begin{equation}
           R(l) \sim l^{\nu_{w}}   .
\end{equation}
The corresponding two-point function is
\begin{equation}
    D(x)=\int_{0}^{\infty} dl e^{-\theta l} P(x,l) .   \label{exx}
\end{equation}
The parameters are associated with physical properties of the walk. The energy
per unit length is $\theta$ \cite{r10}, the typical distance from the starting
point for a walk of length $l$ is $R(l)$,
and the Hausdorff dimension of the walk is $d_{w}=1/\nu_{w}$.
For large $x$,
\begin{equation}
   D(x) \sim e^{-m_{P}x}
\end{equation}
with $m_{P} \sim \theta^{\nu_{w}}$.

The purpose of this paper is to elaborate the physical picture of the walk
representation so that the origin of nontrivial critical exponents can be
understood. Critical behavior of the field theory is controlled by the
deviation of the bare mass $m$ from its critical value $m_{c}$ for which
$m_{P}=0$. A convenient parameter is $t=m^{2}-m_{c}^{2}$. If $\theta
\propto t$, then $m_{P} \sim t^{\nu_{w}}$ and $\nu=\nu_{w}$. However, we
will see that things are more complicated than that. We discuss scalar
field theory in the language of walks and obtain some understanding of the
physical origin of the exponent $\nu$. 

The walk representation is not powerful enough to give a new method for
calculating the exponents. However, it contributes an interpretation of the
calculations carried out by other methods. Alternatively, one may say that
the results of standard field theory calculations can be used to deduce the
properties of the field theory walks. 

At first, one might think that direct self-interactions of the walk that
connects the two points could account for the nontrivial value for $\nu$.
However, this always gives the self-avoiding walk exponent
$\nu=\nu_{saw}\approx0.59$. More structure is needed, and it is present in
field theory. The connecting walk of the two-point function travels through
a background of fluctuations of the field. This can be thought of as a gas
of closed walks. The problem is that when the control parameter $t$ is
adjusted to criticality, it is not just the connecting walk that has
divergent behavior. At the same parameter value, the background in which it
travels has singularities in its properties. The main work of this paper
will be in disentangling these effects. 

The physical mass of the field is $m_{P}$, and
the background sets a distance scale $m_{P}^{-1}$. Three regions can be
considered:
\begin{description}
  \item[R1] $R\gg m_{P}^{-1}$
  \item[R2] $R\simeq m_{P}^{-1}$
  \item[R3] $R\ll m_{P}^{-1}$
\end{description}
The relevant one is R2; however, it is best understood as part of a
larger picture that includes all three regions.
Only in R1 do the direct self-interactions fully account for the
correct value of the exponent in the corresponding region.
In R2 and R3, the effects from the background must also be included.

A free random walk has no self-interaction and no interaction with a
background of loops. The energy--entropy balance gives a nice physical
picture of the second order phase transition, but the correlation length
exponent of 0.5 that follows is correct for a free field only. While the
walk picture and the energy--entropy balance can be retained, interactions
of the walk must be added. A self-avoiding walk has repulsive
self-interactions. (These need not be infinitely repulsive.) The repulsive
interaction occurs within a fixed range. Different elements of the walk
interact if they are within that distance of each other. As the walks
become very long at the critical point, the interaction range does not
grow. This effect causes the shift $0.5\rightarrow\nu_{saw}$. It is
independent of the details of the repulsion (universality). An additional
shift is needed to get the correct value for an interacting field theory.
This must come from a distance scale for the self-interaction that grows
without bound as the critical point is approached. This long-range
self-interaction is induced by the interaction of different elements of the
walk with the background loops, which have a size that diverges at the
critical point. 

The most important observation is that the singular behavior of the background
induces nonanalytic behavior in the relation between the mass parameter
$m^{2}=m_{c}^{2}+t$
and the energy per unit length
\begin{equation}
    \theta \sim t^{\nu_{\theta}}.
\end{equation}
From this, it follows that
\begin{equation}
    \nu=\nu_{\theta} \nu_{w}=\nu_{\theta}/d_{w}.
\end{equation}
The exponent $\nu_{\theta}$ can be calculated and is equal to a crossover
exponent of the field theory. Thus, the physical origin of the correlation
length exponent is two-fold. Important effects come from
background-induced self-interactions of
the walk, which have a diverging range, and from singular background
contributions to the energy per unit length.

Section 2 contains the technical discussion that leads to these claims.
The description in Secs. 2.1--2.3 applies to all three regions. The
specifics of R1, R2, and R3 appear separately in Secs. 2.4.1--2.4.3. 

\section{Analysis}
\setcounter{equation}{0}
The first order of business is to derive the representation of the
two-point function that will be used in the rest of this paper. It is a
particular version of a proper time path integral. The length
interpretation of the proper time will be emphasized. The two-point
function is an integral over the length of the connecting path with a
weight that is a function of the length and will be discussed in detail.
Properties of the weight are associated with geometrical properties of the
walk. From the known behavior of the two-point function, there follows some
specific results for the walk. It is important to separate the effects of
how the end points of walks of a given length are distributed in space from
those of how the weight depends on the length and the bare mass. 

\subsection{Representation}
\subsubsection{Preliminary formalism}

Consider a scalar field that is quadratically coupled to an external potential
$U(x)$.
The two-point function ${\cal D}(x,y;v)$ is the solution to
\begin{equation}
(-\partial^{2}+v(x)){\cal D}(x,y;v)=\delta(x-y)
\end{equation}
with $v(x)=U(x)+m^{2}$. Let $t$ be the deviation of $m^{2}$ from some
preferred value $m_{c}^{2}$, so that $v=U+m_{c}^{2}+t=V(x)+t$. Now
introduce the proper time $l$, which is proportional to the length, and the
function $G(l;x,y;V)$ which solves 
\begin{equation}
  - \frac{\partial}{\partial l}G=HG                \label{e1}
\end{equation}
with $H=-\partial^{2}+V(x)$. Then
\begin{equation}
      {\cal D}=\int_{0}^{\infty} dl e^{-tl}G.                     \label{e2}
\end{equation}
Since the solution to (2.2) is
\begin{equation}
       G=e^{-Hl}             \label{???}
\end{equation}
with
\begin{equation}
       G(0;x,y;V)=\delta(x-y),
\end{equation}
G has a standard path integral representation as a sum over all paths with
$x(0)=y$ and $x(l)=x$. 
\begin{equation}
    G(l;x,y;V)= \prod \int_{x(0)=y,x(l)=x} dx
          e^{-\int_{0}^{l}d\tau[\frac{1}{4}\dot{x}^{2}+V]}
\end{equation}
The length $L$ of these paths (walks) \cite{r11}
is proportional to $l$ and inversely proportional to the spatial cutoff
$a$ \cite{r4}.

\subsubsection{O(N) field theory}
At fixed spatial cutoff $a=1 / \Lambda$, the $O(N)$ symmetric Lagrangian is
\begin{equation}
 {\cal L}=\frac{1}{2} (\partial \phi)^{2}+\frac{1}{2} m^{2} \phi^{2}
          +\frac{g}{4!} \phi^{4}
\end{equation}
The bare mass squared $m^{2}$ is
\begin{equation}
  m^{2}=m_{c}^{2}+t.   \label{e3}
\end{equation}
The critical bare mass squared $m_{c}^{2}$ is a function of the cutoff and the
bare coupling $g$ chosen so that at $t=0$ the physical mass $m_{P}$ is zero.

The usual device of writing
\begin{equation}
  e^{-\frac{g}{4!} (\phi^{2})^{2}}=
   \frac{1}{\surd(2\pi)} \int d\sigma
     e^{-\frac{1}{2}\sigma^{2}+b\sigma \phi^{2}}
\end{equation}
with
\begin{equation}
    b=i\surd(\frac{2g}{4!})                           ,
\end{equation}
allows the $\phi$ integral to be done. The result for the two-point
function is
\begin{equation}
  <\phi_{A}(x)\phi_{B}(y)>=\delta_{A,B}D(x-y)
\end{equation}
with
\begin{equation}
    D(x-y)=Z^{-1} \prod \int d\sigma e^{-\frac{1}{2}\int dx \sigma^{2}
       -\frac{N}{2}Tr \ln [-\partial^{2}+m^{2}-2 b \sigma]}
                {\cal D}(x,y;m^{2}-2 b \sigma)  .
\end{equation}
Inserting (2.3) gives
\begin{equation}
  D(x-y)=\int dl e^{-tl}W
\end{equation}
with
\begin{eqnarray}
 W & = & Z^{-1} \prod \int d\sigma
         e^{-\frac{1}{2}\int dx \sigma^{2}
            -\frac{N}{2} Tr \ln [-\partial^{2}+m^{2}-2 b \sigma]}
                 G(l;x,y;m_{c}^{2}-2 b \sigma)  \nonumber \\
   & = & \prod \int_{x(0)=y,x(l)=x} dx \prod \int d\sigma
         e^{-\int_{0}^{l}d\tau[\frac{1}{4}
                 \dot{x}^{2}+m_{c}^{2}-2 b \sigma]} \nonumber \\
   &   &     \times Z^{-1}e^{-\frac{1}{2}\int dx \sigma^{2}
                  -\frac{N}{2}Tr \ln [-\partial^{2}+m^{2}-2 b \sigma]} .
\end{eqnarray}

The problem with this representation is that $W$ depends on $t$ so that the
dependence of $D$ on $t$ cannot be used to determine $\nu_{w}$. The $t$
dependence in $e^{-tl}$ and $W$ can be disentangled by the following
device: First, extend the model to $O(N+M)$ symmetry by introducing another
$M$ fields called $\chi$. The new Lagrangian is 
\begin{equation}
 {\cal L}=\frac{1}{2} (\partial \phi)^{2}+\frac{1}{2}m^{2}\phi^{2}
         + \frac{1}{2} (\partial \chi)^{2}+\frac{1}{2}m^{2}\chi^{2}
           +\frac{g}{4!} (\phi^{2}+\chi^{2})^{2}  .
\end{equation}
Second, softly break the $O(N+M)$ symmetry to $O(N)\times O(M)$ with a
different bare mass for the $\chi$ field. 
\begin{equation}
  m^{2}(\phi^{2}+\chi^{2})\rightarrow m^{2}\phi^{2}+m'^{2}\chi^{2}
\end{equation}
Now in addition to (2.8),
\begin{equation}
   m'^{2}=m_{c}^{2}+t' ,
\end{equation}
and $m_{c}$ is the correct function of $\Lambda$ and $g$ so that at $t=t'=0$,
$m'_{P}=m_{P}=0$.

The two-point function for the first component $\chi$ is
\begin{eqnarray}
  D'(x-y)&=&<\chi_{1}(x)\chi_{1}(y)>  \nonumber \\
         &=&\int dl e^{-t'l}W'  \nonumber \\
         &=&\int dl e^{-t'l} Z^{-1} \prod \int d\sigma
             e^{-\frac{1}{2}\int dx \sigma^{2}
       -\frac{N}{2}Tr \ln [-\partial^{2}+m^{2}-2 b \sigma]
       -\frac{M}{2}Tr \ln [-\partial^{2}+m'^{2}-2 b \sigma]}  \nonumber \\
         & & \times G(l;x,y;m_{c}^{2}-2 b \sigma)  .
                 \label{e4}
\end{eqnarray}
Evidently,
\begin{equation}
   D'|_{M=0,t'=t}=D.
\end{equation}
It is important to note that in (2.18) the weight $W'$ is independent of
$t'$ and equal to $W$ when $M=0$.

\subsection{Properties of the weight $W'$}

The weight $W'$ is a function of the variables $x,l,t,M,N,g,$ and
$\Lambda$. In this discussion, $g,N$, and $\Lambda$ are fixed. Then $t$
determines the physical mass $m_{P}$ with $m_{P}=0$ for $M=0$ and $t=0$.
Since only $m_{P}^{-1},x$ and $l^{\frac{1}{2}} \gg a$ are of interest, the
important thing is relations among $m_{P},x$, and $l$. The variables $g,N$,
and $\Lambda$ can be ignored. 

Consider the large $l$ behavior of $W'$. To anticipate some of the results,
observe that $W'$ can be interpreted as the exponential of the excess free
energy of the field due to the presence of the path. This interpretation
suggests the presence of a factor 
\begin{equation}
  e^{-F}
\end{equation}
with
\begin{equation}
    F=fl .  \label{e5}
\end{equation}
If the path has dimension $d_{w}$, then a singular piece in $f$ of the form
\begin{equation}
    f_{s} \propto m_{P}^{d_{w}}  \label{eyy}
\end{equation}
might be expected.

For a more precise treatment, define
\begin{equation}
   t_{+}=\frac{Mt'+Nt}{N+M}
\end{equation}
\begin{equation}
   t_{-}=\frac{t'-t}{N+M} .
\end{equation}
The inverse relations are
\begin{equation}
  t=t_{+}-Mt_{-}
\end{equation}
\begin{equation}
 t'=t_{+}+Nt_{-}  .
\end{equation}
The discussion in Ref.\cite{r5} shows that the critical curve for the $\chi$
field is
\begin{equation}
   t_{-} \propto t_{+}^{\varphi}
\end{equation}
or for the case at hand with $M\rightarrow 0$
\begin{equation}
  t'_{c}=t+ct^{\varphi}  .        \label{e6}
\end{equation}
The crossover exponent
\begin{equation}
    \varphi\approx 1+\frac{N}{2(N+8)} \epsilon .
\end{equation}
Equivalently,
\begin{equation}
   m'_{P}=m_{P} {\cal F}(t_{-}/t_{+}^{\varphi}) .      \label{e14}
\end{equation}

Now return to the form of $F$ in (2.21). The complete exponent is
\begin{equation}
   t'l+F
\end{equation}
and by assumption the factor
\begin{equation}
   e^{-(t'l+F)}    \label{e7}
\end{equation}
contains the dominant large $l$ behavior of the integrand for $D$. $F$ does
not depend on $t'$. The second order phase transition occurs when this
exponential factor no longer damps the $l$ integral. The position of the
critical point (2.28) is independent of $x$. Thus, the factor
(2.32) must loose its large $l$ decay when $t'=t'_{c}$, and this can
happen only if $F\propto l$ for large $l$ as in (2.21). 

Now it follows that
\begin{equation}
    t'_{c}+f(t)=0 .
\end{equation}
With (2.28), this gives
\begin{equation}
    f=-(t+ct^{\varphi})
\end{equation}
and an integrand proportional to
\begin{equation}
     e^{-\theta l}
\end{equation}
with
\begin{equation}
   \theta=t'-t-ct^{\varphi}=t'-t'_{c}.    \label{e20}
\end{equation}

With the introduction of $P$ through
\begin{equation}
   e^{-lt'}W'=e^{-\theta l}P ,
\end{equation}
the representation (2.18) for the field becomes
\begin{equation}
    D'(x-y)=\int dl e^{-\theta l}P(x,l,t) ,   \label{e10}
\end{equation}
which is similar to (1.3). By definition, $P$ does not contain any
factors with the exponential $l$ dependence of the first factor of the
integrand. The energy per unit length of the path is $\theta$. It is
important to note that $\theta$ is proportional to $t^{\varphi}$ not $t$
when $t'=t$. This enters through $F$ and is due to the interactions of the
path with the background. 

\subsection{Scaling forms}

The next topic is the scaling behavior of $P$. This can be deduced from the
known behavior of $D'$, which is
\begin{equation}
    D'(x)=(m'_{P})^{q} C(xm'_{P})
\end{equation}
with $q=d-\gamma ' / \nu '$ and
\begin{equation}
    m'_{P}\propto (t'-t'_{c})^{\nu'} .
\end{equation}
Now use the representation (2.38) and assume a form for $\theta$ that is
slightly more general than (2.36).
\begin{equation}
  \theta=(t'-t'_{c})^{\nu_{\theta}}
\end{equation}
Since $P$ is the inverse Laplace
transform of $D'$, we get the scaling form of $P$.
\begin{equation}
    P=p(x/R(l))        \label{e81}
\end{equation}
with
\begin{equation}
   R\propto l^{\nu_{w}}    \label{e8}
\end{equation}
and
\begin{equation}
    \nu_{w}=\nu' / \nu_{\theta}   \label{e9}
\end{equation}
while
\begin{equation}
   d_{w}=1/ \nu_{w} = \nu_{\theta} / \nu '  .   \label{e23}
\end{equation}

The exponential in (2.38) limits $l$ to
\begin{equation}
    l \alt l_{M}=\theta^{-1}  .
\end{equation}
Quantities of the same order are
\begin{equation}
    m'_{P} \sim [R(l_{M})]^{-1} \sim l_{M}^{-\nu_{w}} \sim \theta^{\nu_{w}} .
                      \label{e11}
\end{equation}

\subsection{Three regions}

A complete description of the properties of the paths of the field theory
two-point function must distinguish the three regions R1, R2, and R3. With
$M=0$, the $t$ dependence of $P$ determines a scale $m_{P}^{-1} \sim
t^{-\nu}$, which comes from the background field $\phi$. Paths for which
$R$ is much greater than or much less than $m_{P}^{-1}$ have different
properties. Fortunately, we now have some control over this through the
ability to adjust $x$ and $l$ without changing $m_{P}^{-1}$. As will be
seen, the dimension of the paths $d_{w}$ has one value for $x$ and $R$ $\gg
m_{P}^{-1}$ and another for $x$ and $R$ $\alt m_{P}^{-1}$. 

\subsubsection{R1}

This is the region with $R \gg m_{P}^{-1}$. In it, $m_{P}$ is fixed and
finite while $m'_{P} \rightarrow 0$. To obtain this, keep $t$ finite and
send $t' \rightarrow t'_{c}$. Thus $\theta \rightarrow 0$ with
$\nu_{\theta}=1$, $m'_{P} \rightarrow 0$, and $l_{M} \rightarrow \infty$.
Choose $x \sim (m'_{P})^{-1}$ so that $x, (m'_{P})^{-1}$, and $R(l)$ are
all much larger than $m_{P}^{-1}$. 

This is a critical $O(M)$ theory with $M \rightarrow 0$.
It is the
field theory model of the self-avoiding walk with
$\nu'=\nu_{saw} \approx 0.59$ \cite{r6}. It follows that
\begin{equation}
    d_{w}=1/\nu_{w}=1/\nu_{saw} \approx 1.69  .  \label{e13}
\end{equation}
This analysis shows that (2.48) is a generic result for paths with any
sort of self-repulsion that acts over a fixed, finite range.
As mentioned already, $\nu_{saw}$ is not the correct value for $\nu$ in
the $O(N)$ theory.

\subsubsection{R3}

The opposite extreme has $x$ and $R$ $\ll m_{P}^{-1}\rightarrow \infty$. To
study this, set $M=0$ and take $t \rightarrow 0$ so that
$m_{P}^{-1}\sim t^{-\nu} \rightarrow \infty$ while $m_{P}^{'-1}$ stays
finite. It follows from the scaling form (2.30) that
${\cal F}(z)$ must behave as $z^{\nu / \varphi}$ at large $z$ and that
\begin{equation}
    m'_{P} \sim t'^{\nu / \varphi} .       \label{ep1}
\end{equation}
Also $t'_{c}=0$, and $\theta=t'$, so $\nu_{\theta}=1$.
To get $\nu'$, send $t' \rightarrow t'_{c} = 0$ so that
$m_{P}^{'-1} \rightarrow \infty$. Equation (2.49) says that
$\nu'= \nu / \varphi$.
Since $\nu_{w}=\nu' / \nu_{\theta}$,
\begin{equation}
    d_{w}=1/\nu_{w}=\varphi/\nu.
\end{equation}

Now it is evident that indirect self-interactions of the path, which are
induced by the background and {\em act on arbitrarily large scales}, cause
an additional change in the dimension of the path. For small $\epsilon$,
this dimension is closer to 2 in R3 than it is in R1. The indirect
interactions must be an  effective attraction that competes with the
repulsion already discussed. If the path has passed through a certain
volume and repelled the background there, then it will be less costly for
another part of the path to go through that volume --- an effective
attraction weakening the direct repulsion. 

While the exponent $\nu'$ in R3 has a value that is different from that in
R1, it is still not the correct exponent for the $\phi$ field. Further
understanding awaits in a consideration of R2. 

\subsubsection{R2}

Here we want to lie between R1 and R3 and have $m'_{P} \sim m_{P}$. This
includes the original point of interest $m'_{P} = m_{P}$. It follows from
(2.30) that we should take $t' \rightarrow t'_{c}$ with
$t_{-}/t_{+}^{\varphi}$ fixed. For $M=0$, $t'-t \propto t^{\varphi}$, and
$t \sim t'$ since $\varphi > 1$. Thus, $\theta=t'-t-ct^{\varphi} \sim
t^{\varphi} \sim t^{'\varphi}$, $t'_{c}=0$, and $\nu_{\theta}=\varphi$.
With $M=0$, $m_{P} \sim t^{\nu}$. From $t \sim t'$ and $m'_{P} \sim m_{P}$,
it follows that $m'_{P} \sim t^{'\nu}$ and $\nu' =\nu$. Equation
(2.45) shows that this can arise only if 
\begin{equation}
    d_{w}=1/\nu_{w}=\varphi/\nu   .
\end{equation}
This is the same as the result for $d_{w}$ in R3. However, in R2, $\nu'
\neq \nu_{w}=1/d_{w}$ because $\theta$ is proportional to $t'^{\varphi}$
not $t'$. 

Now we see that the dimension of the path in the two-point function is
$d_{w}=\varphi/\nu$ and not $1/\nu$. The correlation length exponent $\nu$
is not $1/d_{w}$ because there is an additional source of nonanalytic
behavior that comes from the singular relation between the bare mass and
the energy per unit length of the path: $\theta \sim t^{\varphi}$ . Also,
notice that $\theta \sim t^{\varphi} \sim (t^{\nu})^{\varphi/\nu} \sim
m_{P}^{d_{w}}$ as suggested in (2.22). 

\section{Conclusion}
\setcounter{equation}{0}

To get $D$ for the original $\phi$ field, use region R2 with $t'=t$ and
$M=0$. As we have shown, 
\begin{equation}
    D(x-y)=\int dl e^{-\theta l}p(x/R(l))
\end{equation}
with
\begin{equation}
   \theta \sim t^{\varphi},      \label{e25}
\end{equation}
\begin{equation}
   R(l) \sim l^{1/d_{w}},
\end{equation}
and
\begin{equation}
    d_{w}=\varphi / \nu.
\end{equation}
The path connecting $x$ and $y$ has a nontrivial dimension satisfying
\begin{equation}
    d_{saw} < d_{w} < 2.
\end{equation}
There is a singular relation (3.2) between the energy per unit length
and the bare mass that is due to the singular behavior of the background at
$t=0$, $m_{P}=0$. 

\section*{Acknowledgements}

This research was supported by the U.S. Department of Energy.

\bibliographystyle{unsrt}

\begin{thebibliography}{9}
\bibitem{r10} In some contexts, $\theta$ is a real energy per unit length
              while in others (such as this) it is only analogous to that.
              We will call it an ``energy'' in any case.
\bibitem{r11} We use the terms ``walk'' and ``path''
              without any technical distinction in meaning. We tend toward
              ``walk'' when we have a lattice cut-off in mind and toward
              ``path'' when thinking of continuous space.
\end{thebibliography}

\end{document}